\documentclass{article}
\pdfoutput=1

\usepackage{arxiv}

\usepackage[utf8]{inputenc} 
\usepackage[T1]{fontenc}    
\usepackage{hyperref}       
\usepackage{url}            
\usepackage{booktabs}       
\usepackage{amsmath}
\usepackage{nicefrac}       
\usepackage{microtype}      
\usepackage{caption}
\usepackage{subcaption}
\usepackage{graphicx}

\title{Detection of symmetry using a crystallographic image processing algorithm.}
\author{Paul Plachinda \\
  Department of Physics\\
  Portland State University\\
  Portland, OR 97201 \\
  \texttt{plachind@pdx.edu}
}
\begin{document}
\maketitle

\begin{abstract}
This article presents an automated method to quantify and detect symmetry elements in 2D
patterns by means of image processing. Escher’s woodcuts, a widely recognized didactic tool for
crystallographic education of students, were used to demonstrate this approach. We also discuss
peculiarities in the detection of black and white symmetry, color symmetry, and detection of the
”hidden” and ”broken” symmetry elements by means of the phase origin map approach.
\end{abstract}

\section{Introduction}
The legacy of the Dutch graphic artist M.C. Escher (1898 - 1972) consists of more than 500 woodcuts, many of which exhibit 
exceptionally peculiar symmetric patterns,\footnote{All M.C. Escher works \copyright 2002 Cordon Art - Baarn - Holland (www.mcescher.com). All rights reserved. Used by Fair Use License} 
which combine outstanding artistic beauty with a carefully elaborated symmetry. These two unique points have made them a valuable tool for teaching the foundations of symmetry in multiple elementary textbooks on crystallography.\cite{schwarzenbach1996crystallography,gould2013crystal} 
Besides tessellations of the plane Escher's drawings demonstrate more complex concepts, such as black and white, and color symmetry. The founders of these concepts, Shubnikov and Belov\cite{shubnikov1974symmetry, shubnikov1964colored} also used Escher's drawings to illustrate their pioneering ideas. The physical application of these concepts is magnetic symmetry, where the black and white groups correspond to spin up/spin down states of the same atom in the unit cell. Color symmetry is used to describe more complex spin arrangements.

Teaching symmetry to students can be sometimes complicated since the mathematical representation of it as thought at the university level sometimes contradicts student's naive perception. Several papers (e.g.\cite{Nussbaum} and \cite{Amezcua}) have previously addressed comprehension difficulties and provided simple models to understand cornerstone concepts of crystallography and symmetry. 

Visual perception of symmetry by humans can be illustrated by several examples:\cite{Landwehr} people tend to value more and pay attention to high-symmetry elements and perfectly repeating patterns, and to ignore features violating such. Violation of translational symmetry is considered exceptionally more obvious and disturbing rather than violation of  rotational or mirror symmetry elements. It is interesting to observe, that traditional Japanese aesthetics (Wabi-sabi) values asymmetry more than symmetric patterns associated with the Greek ideals of beauty and perfection in the West.\cite{koren2008wabi}  

From the author's personal teaching experience, students tend to envision symmetry in two different ways: an orthodox way - if there is a single feature violating proposed symmetry, the overall symmetry is dictated by this feature; and a relaxed way - if there is enough repeating motifs of the same symmetry, then the image possesses the symmetry of an undistorted motif. Real crystals never possess perfect symmetry as described by their mathematical model due to point and line defects, grain boundaries, magnetic domains, etc. This paper is aimed to elucidate the concepts of symmetry quantification, refinement, and detection of the "hidden" and "broken" symmetry elements. 

For more detailed consideration we chose four examples: three Escher's plane tessellations (Figures~\ref{E45_img},\ref{Newts_img},\ref{lizards_img}) and one so called "impossible figure" (Figure ~\ref{Imp_img}). We will discuss the ability of the algorithm developed by Zou and Hovm\"oller\cite{Zou2006} to assist in determination of correct plane group via so called residuals (Examples \ref{Angles} and \ref{Imp}). Also we discuss usefulness of the Phase Origin map (POM) to reveal "broken" (Example \ref{Imp}) and "hidden" (Example \ref{newts} and \ref{lizards}) symmetry elements, often associated with colorization of plane tessellations. 

 \section{Theoretical Methods}
\subsection{Crystallographic image processing}

The essence of Crystallographic image processing (CIP) and its the basic steps outlined in the quote of Sir Aaron Klug, a Nobel Prize winner and pioneer of the CIP\cite{klug}: "The essence of image processing of this type is that it is a two-step procedure after the first image has been obtained.  First the Fourier transform of the raw image is produced.  Next, Fourier coefficients are manipulated, or otherwise corrected, and then transformed back again to reproduce the reconstructed image.” Applicability of the CIP to high resolution TEM (HRTEM) images to facilitate crystal structure determination has been extensively developed in the research group of Sven Hovm\"oller. (See e.g. \cite{Zou2006}) A great amount of real live examples is use of the CIP techniques in application to crystal structure determination from HRTEM can be found in~\cite{ElDiff}.  First application of this technique to images other than HRTEM, namely to images coming from a scanning tunnelling microscope (STM), was done by P.Moeck.\cite{moeck2012systems} Also it had been pointed out that the CIP techniques can aid in removal of scanning artifacts.\cite{Straton2015} 

Here we make the definition of the  CIP agnostic of the method how the images were obtained and whether they depict a crystal structure or not: all images that contain any intrinsic symmetry, including just translational symmetry, can be processed "crystalographically". To demostrated this concept alongside with some numerical peculiarities of the CIP methods we will apply them to Escher's drawings.

This family of the CIP methods is based on analysis of a symmetric pattern in the Fourier (inverse) space. Since the image periodic, its Fourier space representation is build up by a discrete set of complex-valued Fourier components (FC).
The Fourier transformation has direct relation with electron or X-ray diffraction - methods widely used for crystal structure determination. Electrons (or X-rays) diffract on certain crystallographic planes with Miller indexes $(hk)$, or $(hkl)$ in the 3D space, however we limit our scope here to 2D drawings/structures. Therefore, each peak in diffraction picture is labeled by the indexes of the corresponding Miller plane. Similarly each peak in the reciprocal (or Fourier) space is also assigned a pair $(hk)$ indicating its diffraction "order".  
The FCs are also commonly referred in X-ray and electron crystallography as "structure factors" or "reflections", here, however, we refrain from using this term since it mostly applies to experimental methods of structure determination.  

A plane (or space groups) possess a certain number of symmetry elements linking together the atomic positions in the real space. Correspondingly this symmetry is preserved upon the Fourier transformation, and now the magnitudes and the phases of certain FCs in the reciprocal space are not random anymore, but have to match with those of another FC connected to the first one by a symmetry opration. Further details are discussed in section \ref{Symmetrization}.

\subsection{Phase statistics in ALLSPACE and CRISP}
The CIP algorithm is based on certain relations between the phases ($\phi(hk)$) and the amplitudes ($F(hk)$) of the Fourier components (FC) within a given plane or space group. 
Traditional way to compute phase statistics was implemented in the ALLSPACE (2dx, www.2dx.unibas.ch) program, compares FCs which phases should be $0^{\circ}$ or $180^{\circ}$ with their actual phases, and separately compares phases of low order FCs (i.e. for which $h$ and $k$ are small, as defined by the authors of the program) with their symmetry related mates. 
R-factor is then employed as a measure of goodness of fit, defined as 
$R=\frac{\sum \lvert \phi_{obs}-\phi_{calc}\rvert}{\sum \lvert \phi_{obs}\rvert}$. 
Here $\phi_{obs}(hk)$ is the experimentally observed phase, and $\phi_{calc}(hk)$ - the theoretical phase obtained from a proposed structural model. 
Thus, two different R-factors are available in ALLSPACE: "vs. theoretical", i.e. where $\phi_{calc}$ is calculated based solely on the plane group information, and "vs. other spots" where the program attempts to link pair of symmetry related and calculate mutual phase discrepancy. Further details about the ALLSPACE algorithm can be found in the web page.

Another approach proposed by Zou and Hovm\"oller,\cite{Zou2006} is an algorithm for the evaluation of the so called "symmetrized phase", defined as 
\begin{equation}
\label{phi_sym}
\phi_{sym}(hk)=\arctan\left(\frac{\sum_j s^j w^j \sin(\phi_{obs}^j(hk))}{\sum_j s^j w^j \cos(\phi_{obs}^j(hk))}\right)+
\\
\begin{cases}
0 ^{\circ} & \text{ if } \sum_j s^j w^j \cos(\phi_{obs}^j(hk))>0 \\ 
180^{\circ} & \text{ if } \sum_j s^j w^j \cos(\phi_{obs}^j(hk))<0 
\end{cases}
\end{equation}
The summation in Eq.~\eqref{phi_sym} runs over all symmetry-related FCs defined by the desired symmetry group; $w^j$ is a weighting factor, often set to the amplitude of the corresponding FC, $s^j=1$ if the phases $\phi_{sym}(hk)$ and $\phi_{obs}^j(hk)$ should be equal from the plane group relations and $s^j= -1$ if they should differ by $180^{\circ}$. If a FC is not related to any other FC by the symmetry (except by the Friedel's law), then: $\phi_{sym}(hk)= \phi_{obs}(hk)$. Derivation of this formula and a computationally efficient version of it is given in the Appendix.   

Optimal phasing is determined as a minimum of the "phase residual" functional
\begin{equation}
\label{phi_res}
\phi_{Res} \left[\phi_{obs}(hk)\right] =
\sum_{hk} F_{hk}|\phi_{obs}(hk)-\phi_{sym}(hk)|\bigg/\sum_{hk} F_{hk}
\end{equation}
for each plane symmetry group. Here $\phi_{sym}(hk)$ is the "symmetrized phase" which fulfills the symmetry relations and restrictions. 
All information on plane (wallpaper) groups is summarized in the many-volume International Tables for Crystallography.\cite{ITA2002,ITB2008} 
A reader is cardinally referred to this compendium of crystallographic information for space and plane group notation, and a comprehensive list of symmetry operators in each group.  

Explicit relations between the FCs for each plane group are tabulated in e.g. Refs. \cite{zou1997electron} and \cite{sym4030379}.  $\phi_{sym}(hk)$ is the same for all symmetry related FCs. Centrosymmetric plane symmetry groups ($p2, p2mm, p2gm, p2gg, c2mm, p4mm, p4gm, p6mm$) impose further restrictions on possible values of symmetrized phases limiting them to $0^{\circ}$ or $180^{\circ}$ only. 
The lower the residuals $\phi_{Res}$ are for a given plane symmetry group, the higher is the likelihood that this group is the right one for the pattern under investigation. 

Similarly an "Amplitude residual functional" can be introduced: 
\begin{equation}
F_{Res} \left[F_{obs}(hk)\right] =
\sum_{hk} |F_{obs}(hk)-F_{sym}(hk)|\bigg/\sum_{hk} F_{obs}(hk)
\end{equation}
This functional is identical to the R-work factor, widely used in crystallography and structure determination. 
However, this functional is less important due to the fact the Fourier transform (FT) contains more information in the phase space than in the amplitude space.\cite{1456290} E.g. if we "tamper" the Fourier transform of an image and modify it such way that the amplitudes of all FCs are set the same, but the phases are preserved from the original image, the inverse FT will still yield a recognizable image. If, in turn, the amplitudes are left intact, but all phases are set the same, then the reconstructed image is completely scrambled because it is the phase that determines which FC are present or not, and the amplitudes only determine the contribution of each FC.  

Another quantifier to differentiate between plane groups is the ratio $F_o/F_e$ calculated as the total amplitude of all systematically absent FCs (that were nevertheless observed) to the total amplitude of all other FCs.\cite{Zou2006} 
Systematically absent FCs are the FCs whose $F_e$ amplitudes vanish due to presence of glide line symmetry elements in the plane group ($p1g1$, $p11g$, $c1m1$, $c11m$, $p2mg$, $p2gm$, $p2gg$, $c2mm$, $p4gm$).\cite{ITA2002} (A glide reflection is the composition of a reflection in a line and a translation along that line.)

We use full plane group symbols instead of the reduced ones (i.e. $p1g1$ vs. $pg$) to emphasize the orientation of a symmetry element with respect to the image as it is was drawn. An $F_o/F_e$ ratio is thus another quantifier for "degree of matching" of the image under inspection to a certain to a plane group in addition to the amplitude and the phase residuals. If for an experimental image the $F_o/F_e$ ratio is larger than zero for a given plane symmetry group, the image does not perfectly fit the group potentially due to the presence of hidden and broken symmetry elements --- see below.

These formulas are implemented in the CRISP program.\cite{HOVMOLLER1992121} CRISP is a closed-source program with no possibility to interact with the code for further examination of the algorithm, therefore we have developed our own software implementation of the aforementioned scheme in order to be able to explore the symmetry deeper.\cite{CIP2_MM,CIP_IEEE,CIP_MM}

\section{Algorithm and Design}
\subsection{Image Processing}
The image processing procedure begins with a Fast Fourier Transform (FFT) of the input image. Information is collected exclusively from the Friedel-independent (half-plane) part of the reciprocal space. 
Since any real image is neither truly periodic nor infinite, edge effects such as peak streaking may occur. This causes intensity of the peaks to decrease and makes exact detection of their position more challenging. To reduce this effect, the image is multiplied by a window function before the FFT. 2D Hann window, a common windowing technique in the FFT, has successfully eliminated peak streaking.
Since some plane groups possess systematic absences the amplitudes of some peaks will be zero, which can adversely affect the indexing routine later, therefore we add one more auxiliary operation - an auto cross correlation of the power spectrum (squared amplitude part of the FFT). As the result of this routine, all peaks that previously vanished in the power spectrum are now revealed in their correct positions.    
Once all of these corrections are applied, the program performs a peak search on the power spectrum. It should be noted that though the peak search is performed on the auto-correlated power spectrum, intensities of found peaks are collected from the original power spectrum. 

\subsection{Indexing}
As the result of the peak search the program obtain number pairs $\left\lbrace x_i;y_i\right\rbrace$, called "image coordinates" in pixels, which depend on the size of the image (in pixels). Corresponding pairs in the Fourier space $\left\lbrace h_i;k_i\right\rbrace$, are referred to as "reciprocal lattice coordinates", and depend on the basis vectors assigned to the reciprocal lattice. Within a known basis set $(\vec{U},\vec{V})$ image coordinates and lattice coordinates are mutually linked by the relation:
\begin{equation}
\begin{pmatrix}
	h_i \\
	k_i \\
\end{pmatrix}
=
\begin{pmatrix}
	 U_1 & V_1 \\
	U_2 & V_2 \\
\end{pmatrix}^{-1}
\begin{pmatrix}
	x_i \\
	y_i \\
\end{pmatrix}
\end{equation}
Here $U_1$, $U_2$ are the components of the $\vec{U}$ basis vector and $V_1$, $V_2$ are the components of $\vec{V}$, both expressed in pixels. 
This approach follows somewhat the one described in Ref.~\cite{ZENG2007353}, however this time several more refinement steps were  introduced several more to determine the basis vectors. The choice of basis is done by screening through the low-resolution peaks with weighting factors applied to make rectangular/hexagonal lattices more preferable rather than oblique. The basis, which indexes a sufficient number of low-resolution peaks, is further refined by minimizing the value of functional $L$ using a least squares algorithm (LS).
\begin{equation}
L\left[\vec{U},\vec{V}\right]=\sum_i \left|
\begin{pmatrix}
	 U_1 & V_1 \\
	U_2 & V_2 \\
\end{pmatrix}
\begin{bmatrix}
	h_i \\
	k_i \\
\end{bmatrix}
-
\begin{pmatrix}
	x_i \\
	y_i \\
\end{pmatrix}
\right|^2\quad \textrm{,where []-represents rounding to the nearest integer.}
\end{equation}

As long as the basis is refined and all peaks are indexed, the program can compute "lattice parameters" $\left(a^*,b^*\right)$ as $a^*=\sqrt{U_1^2+U_2^2}$, $b^*=\sqrt{V_1^2+V_2^2}$, $a^*b^*\cos \gamma^*=U_1V_1+U_2V_2$. These values can be further refined via a similar LS routine, as a minimum of the functional $Q$:\\  
$Q\left[a^*,b^*\right]=\sum_i\left(
\left(x_i^2+y_i^2\right)-
\left(h_i^2 a^{*2}+k_i^2 b^{*2}+2 h_i k_i a^*b^*\cos\gamma^* \right)
\right)$
. The value of $\gamma^*$ is not refined, since in all groups except $p2$ and $p1$, the indexing of which is trivial, it is constrained either to $90^{\circ}$ or to $60^{\circ}$. The "lattice parameters" are arbitrary numbers, which depend on the processed image's length calibration and resolution. Therefore exact numbers do not have any real physical sense. However if the ratio between $a^*/b^*$ approaches unity and the angle between both vectors is significantly close to $60^{\circ}$ or $90^{\circ}$, the lattice is likely either hexagonal or square respectively.

The program stores lattice coordinates and FCs $(h, k, F, \phi)$ for each indexed peak for further processing. 

\subsection{Symmetrization}
\label{Symmetrization}
Related FCs are fully determined by the plain group's symmetry operators. In vol. B of the "International tables" (See Ref.~\cite{ITB2008}, Table A1.4.4.1. Crystallographic space groups in reciprocal space, p150) all 3D relations are listed. By dropping the $l$ index in the space groups corresponding to the appropriate plain group (e.g. $Pba2$ $\longrightarrow$ $p2gg$) one can obtain phase relations in the plane groups. Friedel-pairs should be dropped too, since they do not carry any additional information. As one can see from table A1.4.4.1 in Ref.~\cite{ITB2008}, certain groups, like $p1m1$, $c1m1$, $p2mm$, $c2mm$ have the same phase relations. Thus, groups $p1m1$/$c1m1$ and $p2mm$/$c2mm$ can be distinguished by systematical absences due to centering only. Systematical absences contribute to $\phi_{res}$ due to the weighting factors $F_{hk}$. In the $p2$ group there are no relations, restrictions only, i.e. each phase should be restricted to $0^{\circ}/180^{\circ}$.
Once all related pairs are determined, Eq.~\eqref{phi_sym} is applied to calculate a symmetrized phase for each group of related FCs.

\subsection{Phase Origin Map}
An important feature first implemented in CRISP is the "origin search and refine procedure". It utilizes the phase origin map (POM), which is being built by plotting the values of $\phi_{Res}$ for the user-selected plane group as a function of $\phi_x$ and $\phi_y$, related to the arguments of the $\phi_{Res}[\phi_{obs}]$ functional as (See Ref.~\cite{ITB2008}, section 1.4.2.5):
\begin{equation}
\phi \prime _{obs}(hk)= \phi_{obs}(hk)+(\phi_x h+\phi_y k)
\label{phi_shift}
\end{equation}
The POM is computed within the range $\left\lbrace\phi_x,\phi_y\right\rbrace=-180^{\circ}...180^{\circ}$. Then CRISP looks for a minimum on the POM and shifts the phase of each FC $(hk)$ by $(\phi^\textrm{min}_xh+\phi^\textrm{min}_yk)$.
Usability of the POM goes beyond a simple minimum search. Cell metrics can be obtained from the POM as well: its apexes and centering may be recognized by strong, well-localized minima. The symmetry of the POM should be in an ideal case the same as the symmetry of the pattern analyzed. Certain features of the POM will also represent hidden or broken symmetry elements. (See below in the examples section)

\subparagraph{Groups with glide lines}
Some groups ($p1g1$, $p11g$, $p2mg$, $p2gm$, $p2gg$, $p4gm$) exhibit non-trivial phase relations, involving phase shift by a "glide factor" $k\times180^{\circ}$ ($p1g1$, $p2mg$) or $(h+k)\times180^{\circ}$ ($p2gg$, $p4gm$). These factors make estimation of the phase error using just functional \eqref{phi_res} and formula \eqref{phi_sym} inapplicable for groups with glide lines. Therefore, we redefine $s_{hk}$ in formula \eqref{phi_sym} as follows: 
\begin{equation}
s_{hk}=
\begin{cases}
-1 & \text{ if "glide factor" is odd} \\ 
+1 & \text{ if "glide factor" is even} 
\end{cases}
\left(\text{since} \frac{\sin}{\cos}(\phi+k\times180^{\circ})=(-1)^k\times\frac{\sin}{\cos}(\phi)\right)
\end{equation}

Similar correction factors should be applied to functional \eqref{phi_res} by multiplying $\phi_{sym}$ by the $s$-factor: 
\begin{equation}
\label{phi_res_mod}
\phi_{Res} \left[\phi_{obs}(hk)\right] =
\sum_{hk} F_{hk}|\phi_{obs}(hk)-s_{hk}\phi_{sym}(hk)|\bigg/\sum_{hk} F_{hk}
\end{equation}
These corrections guarantee an artifact-free phase origin map with minima corresponding exactly to their real position as determined by the plane group. After shifting phases to these minima as required by Eq.~\eqref{phi_shift} the reconstructed image is expected to be free from symmetry artifacts too.

\subsection{Symmetry Enforcement and Refinement}

"Symmetry enforcement" means the following: For a given plane group, only the Fourier components from the fundamental domain are selected and then the amplitudes and phases from this domain are multiplied by the corresponding symmetry operators onto the whole plane. From this an image is synthesized as a Fourier sum: $I(x,y)=\sum_{h,k}F_{hk}\exp(i\phi_{hk})\exp(i(hx+ky))$. The resultant image is called "enforced" in a certain group.

"Symmetry Refinement" means: A new set of FCs is built, based upon the input data but with symmetry-related FCs set to their respective mean values: phases restricted to either the symmetrized phase or to the nearest $0^{\circ}$ or $+180^{\circ}$ in centrosymmetric groups, phases are shifted according to the POM minimum, or any desired offset. This new set of coefficients is then used to build a new real-space image using the Fourier sum. Fourier sum is a continuous function, that can be used to synthesize image of any pixel size. 
	Enforcement and/or refinement can be performed for any of the plane groups, regardless of the residuals. The residuals are just a rank of the plane groups in terms of their likelihood of matching the original symmetry. Enforced and refined images should be compared with the original ones to ensure that enforcement and refinement was conducted in the correct plane group. 
\section{Discussion}

\subsection{Example 1: "Angel-Devil"}\label{Angles}
\begin{figure}[h!]
\centering
\begin{subfigure}[b]{0.45\textwidth}
    \centering
    \includegraphics[width=\textwidth]{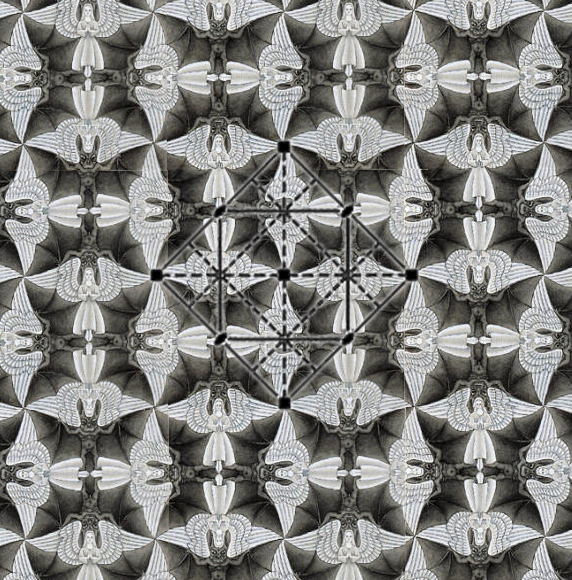}
    \caption{}
    \label{E45_img}
\end{subfigure}
\hfill
\begin{subfigure}[b]{0.45\textwidth}
\centering
\includegraphics[width=\textwidth]{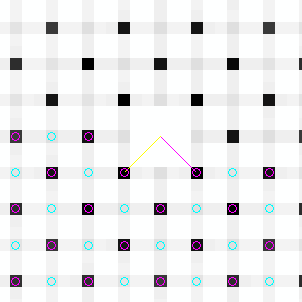}
        \caption{}
        \label{E45_FTT}
\end{subfigure}
\caption{ (a) "Angel-Devil" (No.45), M.C.Escher, 1941, with a $p4gm$ unit cell superimposed. (b) Indexed FFT power spectrum of Figure \ref{E45_img} (central part)}
\end{figure}

As an example of automatic symmetry detection using phase residuals we will consider Escher's woodcut "Angel-Devil". The pattern as follows from the overlaid unit cell (Figure ~\ref{E45_img}), belongs to the $p4mm$ group. This image demonstrates only one concept: the ability of the program to detect the most probable symmetry by analyzing the residuals. Hereinafter "probable group" is assumed in the meaning of the groups having the lowest amplitude and phase residuals. There are other criteria that can go hand in hand with Fourier methods, e.g. Akaike criteria.\cite{sym10050133} 
This image is chosen because angels and devils are considerably different geometrically and thus no "hidden" symmetry elements can connect them. 
After performing FFT of the original image (Figure ~\ref{E45_img}), auto cross correlation, and peak search Figure ~\ref{E45_FTT} is  obtained. The unit cell in the reciprocal space is marked by the blue and green vectors (online only); green is the x-axis translation vector and blue is the y-axis. 

Only the Friedel independent part of the space is used for the peak search. The size of the red circles represents the intensity of the actual Fourier transformation, whereas the white dots correspond to the auto cross correlated image. (Note how FC the previously invisible due to the $g$-glide line are now revealed). 
This pattern can be indexed in a square cell with lattice parameters $a^*=4.0006$, $b^*=3.9992$, $\gamma^*=90^{\circ}$. Slight deviation from the exact relation $a^*=b^*$ is due to accumulation of numerical errors and limited number of repeating elements. Green circles indicate that a peak is indexed in the given basis vector set. The peak search procedure had collected 8264 FCs, within the range of $h_\textrm{max}<99$, $|k_\textrm{max}|<118$. On the next step the program searches related FCs, using the relations as extracted from the table in Ref.~\cite{ITB2008} and computes residuals for a user-selected plane group.

Amplitude and phase residuals for different plane groups are summarized in Table \ref{E45_tab}.

\begin{table}[h!]
\centering
\caption{Amplitude and Phase residuals for Figure \ref{E45_img}}
\begin{tabular}{l c c c}
\toprule
 \textbf{Plane Group} & \textbf{Amplitude Residuals} & \textbf{Phase Residuals} & $F_o/F_e$\\
\midrule	
p2 	& 15.0 & 24.2 & - \\
p1m1 	& 15.7 & 31.0 & - \\
p11m 	& 15.7 & 31.0 & - \\
$p1g1$ 	& 16.0 & 24.1 & 0.087 \\
$p11g$ 	& 16.0 & 24.1 & 0.065 \\
p2mm 	& 16.2 & 53.7 & - \\
$p2gg$ 	& 16.0 & 40.6 & 0.071 \\
$p4$ 	& 14.0 & 41.0 & - \\
p4mm 	& 17.4 & 58.8 & - \\
$p4gm$ 	& 17.5 & 46.5 & 0.071 \\
\bottomrule
\end{tabular}
\label{E45_tab}
\end{table}				  
From Table~\ref{E45_tab}, one can see that the residuals of the italicized groups are lower than the residuals of other klassengleiche groups.\cite{Gramlich2004} In general we look for the highest symmetry possible; therefore we have to compare the image under inspection with the image refined in a certain group. If we refine the image in Figure ~\ref{E45_img} in the $p4gm$ group as well any in any other italicized group from Table~\ref{E45_tab} we observe that the original image doesn't change. Indeed, all italicized groups are the subgroups of $p4gm$. The generating group ($p4gm$) has slightly higher residuals than all its subgroups. This is because the supergroup also has a higher number of FCs mutually connected by symmetry operators. It is clearly seen that groups like $p4mm$ and its subgroups have the residuals higher.

Therefore, formally speaking, every italicized group is a group of symmetry of the image in Figure ~\ref{E45_img}, but $p4gm$ has the highest symmetry of all possible groups. Therefore, from the analysis of the residuals and refined images we can conclude about the plane symmetry of the image under inspection. The image refined in the $p4gm$ group has less noise and image processing artifacts due to averaging multiple FCs by symmetry operators. 

The image in Figure ~\ref{E45_img} has two elements of motif that are clearly distinct from each other. Therefore it does not contain any "hidden" or "broken" symmetry elements. Now we will consider an image with such elements present. 

\subsection{Example 2: "Impossible Figure"}\label{Imp}

\begin{figure}[h!]
\centering
\begin{subfigure}[b]{0.45\textwidth}
    \centering
    \includegraphics[width=\textwidth]{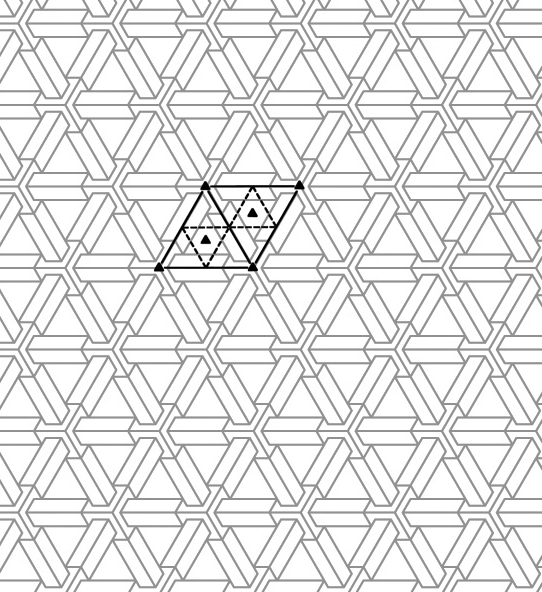}
        \caption{}
        \label{Imp_img}
\end{subfigure}
\hfill
\begin{subfigure}[b]{0.45\textwidth}
\centering
\includegraphics[width=\textwidth]{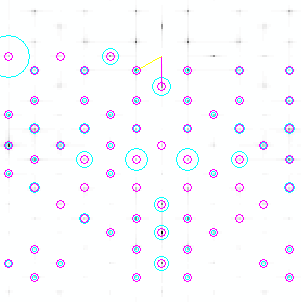}
        \caption{}
        \label{Imp_FTT}
\end{subfigure}
\caption{(a)"Impossible figure" a hexagonal pattern with a $p31m$ group unit cell superimposed. (b) FFT power spectrum of Figure \ref{Imp_img}. FC from the lower semi-plane only are shown}
\end{figure}

In this paragraph we will consider an Escher's inspired hexagonal tessellation on the plane. This is one of so called "Impossible figures".\cite{emmer2007m} The pattern as follows from Figure ~\ref{Imp_img}, belongs to the $p31m$ group. This image is considerably more complex due to presence of elements closely resembling each other but yet not connected by the exact ("obvious") symmetry elements. 
This pattern can be indexed in slightly distorted hexagonal cell with lattice parameters $a^*=14.7559$, $b^*=14.5583$, $\gamma^*=61.38^{\circ}$. Green circles indicate that a peak is indexed in the given basis vector set. The diameter of each red circle indicates initial, not enhanced, amplitude. The peak search procedure had collected 1680 peaks, within the range $|h_\textrm{max}|<36$, $|k_\textrm{max}|<44$

On the next step the program searches for related FCs, using the relations extracted from the table in Ref.~ \cite{ITB2008} and computes residuals for all trigonal/hexagonal ($p3$, $p3m1$, $p31m$, $p6$, $p6mm$) and $p2$ groups.

\begin{figure}[h!]
\centering
\includegraphics[width=0.95\textwidth]{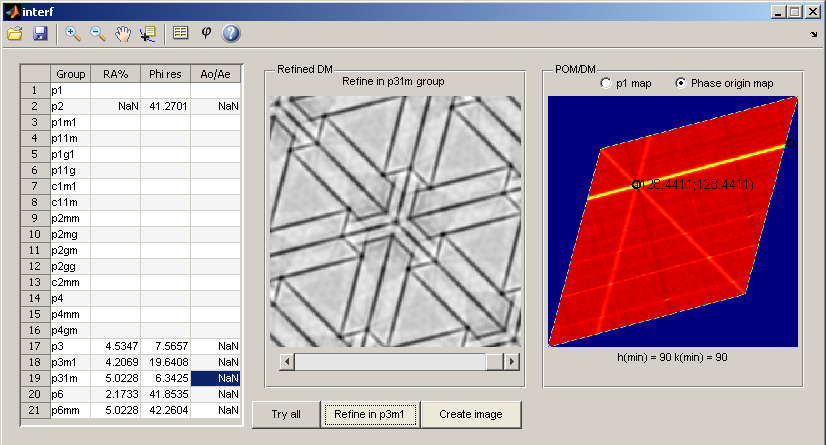}
        \caption{(Program Screen shot) Figure \ref{Imp_img} refined in the $p31m$ group. Positions of the 3-fold axes on the POM are marked with arrows}
        \label{Imp_scr}
\end{figure}		

The phase and amplitude residuals are listed in the table in Figure ~\ref{Imp_scr}. The $p31m$ group has significantly lower phase residuals than all of the other groups. Another group with low phase residuals, $p3$ is a subgroup of $p31m$. The reason group $p3$ has slightly higher residuals, possessing exactly the same symmetry is that there are more symmetry relations taken into account by formula \eqref{phi_sym}.
The image, synthesized using symmetrized phases and $p31m$ group symmetry constrains repeats the input image.

The picture in the right side of Figure ~\ref{Imp_scr} is the Phase Origin Map (POM). It shows $\phi_{Res}$ as a function of $\left\lbrace\phi_x,\phi_y\right\rbrace$. The values of $\left\lbrace\phi^\textrm{min}_x,\phi^\textrm{min}_y\right\rbrace$ corresponding to the lowest $\phi_{Res}$ can be converted into the position of unit cell origin. In Figure ~\ref{Imp_scr} converted position of the minimum is indicated by a circle. A strong line in the POM corresponds to the $m$-line, whereas two dimmer lines intersecting the strong line are, as follows from the unit cell geometry, the $g$-lines. This map shows mutual arrangement of the symmetry elements within the unit cell. The POM with symmetry elements, superimposed on top of it is shown in Figure ~\ref{Imp_img}.

\begin{figure}[h!]
\centering
\includegraphics[width=0.45\textwidth]{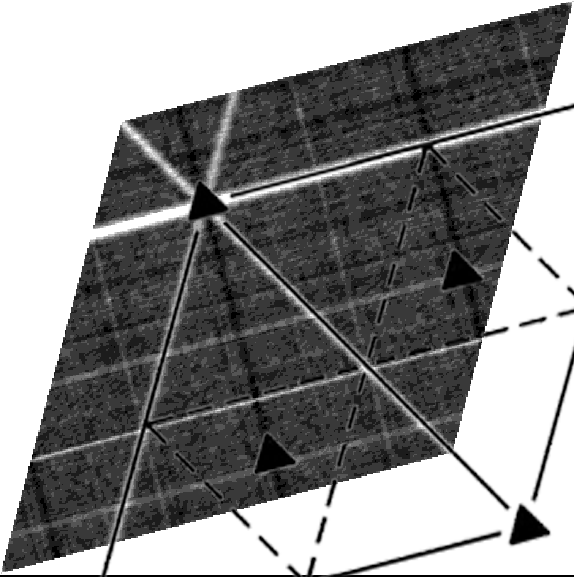}
\caption{POM for the $p31m$ group}
\label{Imp_POM}
\end{figure}		

Black lines in Figure ~\ref{Imp_POM} correspond to a "might-have-been" symmetry element – a vertical mirror line. Only a few features in Figure ~\ref{Imp_img} violate this symmetry element, making it "broken" - this is exactly what makes this image "impossible". The "broken" symmetry group would be $p3m1$. This group also has relatively small residuals. Groups with six fold axes exhibit significantly higher residuals, thus they are not "broken", but rather incorrect. Dark lines in the POM correspond to the rectangular setting of the trigonal lattice.

We would like to emphasize importance of the POM for further crystallographic analysis: By careful reviewing of the POM one can reveal broken or missing symmetry elements. Or symmetry elements which reduce the possible (apparent) high symmetry to lower by exclusion of some symmetry operators due to coloring, defects, twinning, grain boundaries, magnetization, etc. 

\subsection{Example 3:"Lizards"}\label{newts}
This example is devoted to the problem of detection and quantification of black and white symmetry by means of the CIP. The concept of black and white symmetry expands the traditional concept of symmetry by adding a color inversion operator (') to other purely geometrical operators.\cite{shubnikov1974symmetry} Black-and-white groups possess a grey group (i.e. a group where black and white are averaged to grey) as their super group. Strictly speaking, grey groups are not defined for patterns with more than two colors; however, here we will use the concept of "colorblind" images, i.e. images where all the colors are ignored and boundaries are enhanced using the Sobel Filter.

\begin{figure}[h!]
\centering
\includegraphics[width=0.45\textwidth]{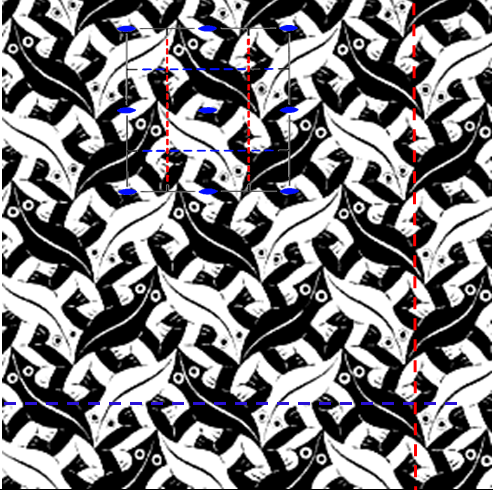}
        \caption{"Lizards" (No.124), M. C. Escher, 1965. $g_y$-lines are shown in red, $g'_x$- in blue, the unit cell of the b/w group $p2'g'g$ is shown as well.}
        \label{Newts_img}
\end{figure}

As an example of the ability of the algorithm to distinguish the black and white symmetry, we will consider Escher's "Lizards" (Figure ~\ref{Newts_img}). By inspection we can determine only one pure geometrical symmetry operator acting on this image: glide lines (dashed red) running though the "cheeks" of the white and the black lizards. Since no other geometrical elements appear, the plane group seems to be $pg_y~(p11g)$ polar.

Initial image processing is done the same way as before. 
Because of use of the autocorrelation during the peak search routine, even peaks corresponding to FCs forbidden by the g-line boost their intensity (e.g. (0,1)). This significantly aids the indexing procedure. The pattern can be indexed in the rectangular system with the lattice parameters $a^*=2.9754$, $b^*=3.0056$. 

\begin{figure}[h]
\centering
\includegraphics[width=0.95\textwidth]{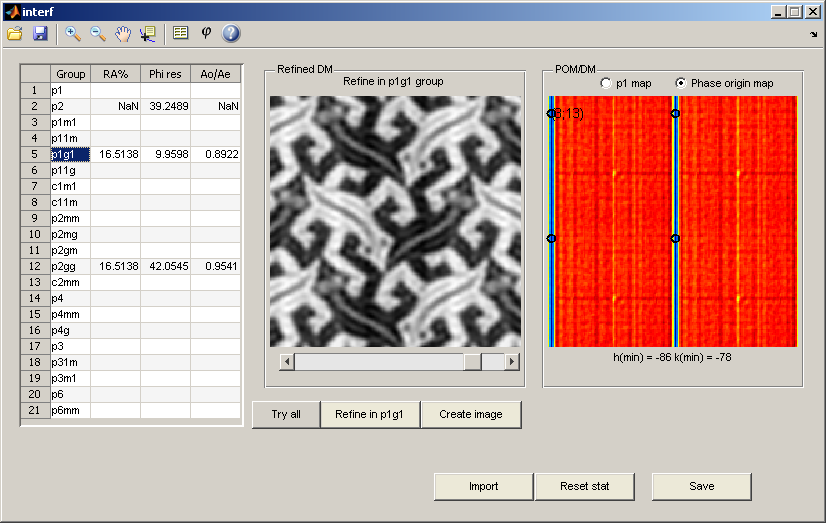}
        \caption{(Program Screen shot) Figure ~\ref{Newts_img} refined in the $p1g1$ group}
        \label{Newts_p1g1}
\end{figure}

On the next step we reconstruct the image and compute phase and amplitude residuals. Since the lizards have very sharp boundaries, FFT produces high number of Fourier components, only a fraction (562) of which was collected.
Group $p1g1$ ($pg_y$) has phase residuals significantly lower than other groups and the reconstructed image perfectly matches the original one. Note the much higher residuals for plane groups $p2$ and $p2gg$. The picture in the right side of Figure ~\ref{Newts_p1g1} is again the POM. Since group $p1g1$ is polar, i.e. has no fixed origin along the $y$ direction, the phase residuals are almost the same along the blue lines. In reality, they differ a little bit due to presence of black-and-white symmetry. Careful examination of the POM in Figure ~\ref{Newts_p1g1} reveals additional highly symmetric features such as yellow and dark red lines, and sharp yellow points. As it was shown before this suggests presence of "hidden" symmetry elements. 

\begin{figure}[h]
\centering
\begin{subfigure}[b]{0.35\textwidth}
    \centering
    \includegraphics[width=\textwidth]{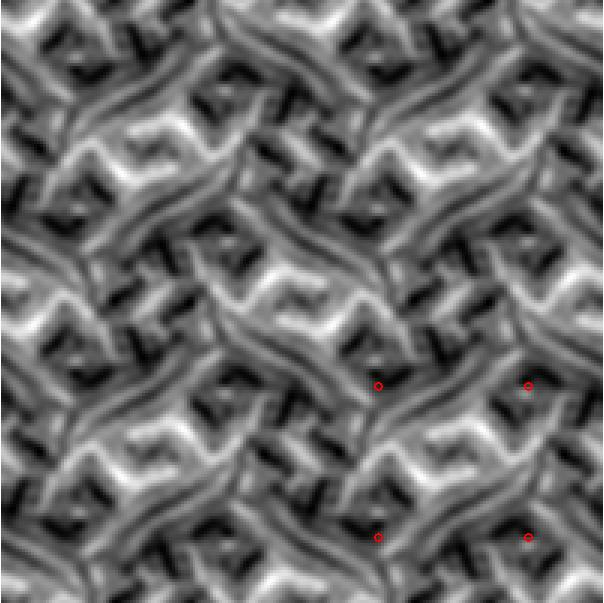}
        \caption{}
        \label{newts_grey}
\end{subfigure}
\hfill
\begin{subfigure}[b]{0.35\textwidth}
\centering
\includegraphics[width=\textwidth]{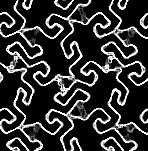}
        \caption{}
        \label{newts_sih}
\end{subfigure}
\caption{(a) "Gray image" refined in $p2gg$ group. (b) Lizards silhouettes (colorblind image)}
\end{figure}

In this case, the "hidden elements" are the elements that change the color of the lizards along with some geometrical manipulations on them. 
Indeed if we refine in a super group of $p1g1$, namely $p2gg$ we obtain a "gray-averaged" image. (Figure ~\ref{newts_grey}). I.e. plane group $p2gg$ is the grey super group for the lizards pattern. 
In Figure \ref{newts_grey} one can still, recognize the lizards' pattern. If black-white symmetry is taken into consideration, one immediately identifies $2'$ axis and $g'_x$ glide lines (depicted in blue in Figure \ref{Newts_img}). Combination of these two symmetry operators with a gray $g_y$ glide line yields the Shubnikov (black and white) group of the image in Figure~\ref{Newts_img}: $p2'gg'$.

\begin{figure}[h!]
\centering
\includegraphics[width=0.95\textwidth]{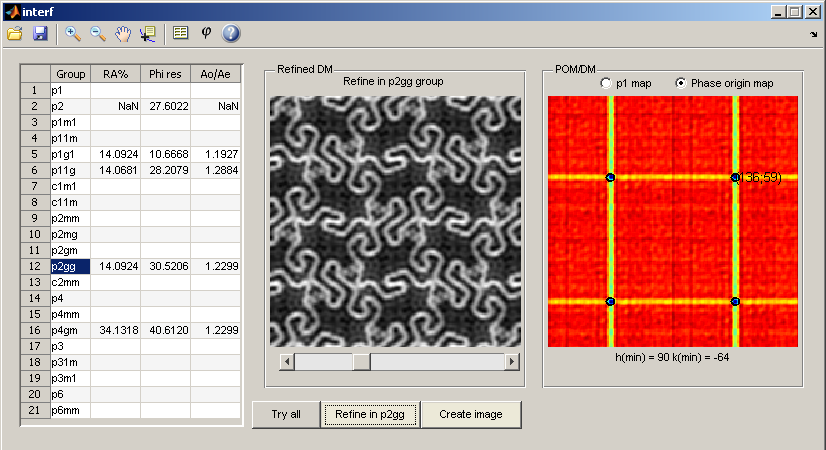}
        \caption{(Program Screen shot) Figure ~\ref{newts_sih} refined in $p2gg$}
        \label{Newts_p2gg}
\end{figure}

We can further investigate symmetry of this image by applying the Sobel filter to Figure ~\ref{Newts_img}. The Sobel filter enhances boundaries and effectively eliminates uniform features (such as colorization), thus yielding a "silhouette" image, where the colors are ignored (See Figure \ref{newts_sih}). In this case the Fedorov (purely geometrical) plane group $p2gg$ is not "broken" anymore and symmetry refinement in this group yields a sharp image. (Figure \ref{Newts_p2gg})

Again, the values of $\left\lbrace\phi_x,\phi_y\right\rbrace$ corresponding to the lowest $\phi_{Res}$ (Figure ~\ref{Newts_p1g1} blue lines, Figure ~\ref{Newts_p2gg} green and yellow lines) can be converted into the position of the unit cell origin(s). In Figure ~\ref{newts_grey} converted positions of these minima are plotted as red rings. They hull the fundamental domain of the pattern. The minima in the POM represent arrangement of the symmetry elements within the cell. In Figure ~\ref{Newts_p2gg} vertical lines are more pronounced than the horizontal, which indicates that the $g_y$ glide-line is less broken than $g_x$, the reason for that is the situation with black and white lizards' eyes (see Figure ~\ref{newts_sih}). While there is a slight difference in the lizards' eyes in Figure ~\ref{newts_sih}, all eyes are averaged out after refinement in the $p2gg$ group. See Figure ~\ref{Newts_p2gg}.
Since only a minor feature of the picture violates the symmetry of the colorblind $p2gg$ group the residuals for $p2gg$ are now closer to the "proper" symmetry group – $p1g1$. Therefore, comparing those two, we can estimate "how much" symmetry is broken by comparing the residuals between "proper" and colorblind groups.

\subsection{Example 4: "Reptiles"}\label{lizards}

\begin{figure}[h!]
\centering
\includegraphics[width=0.45\textwidth]{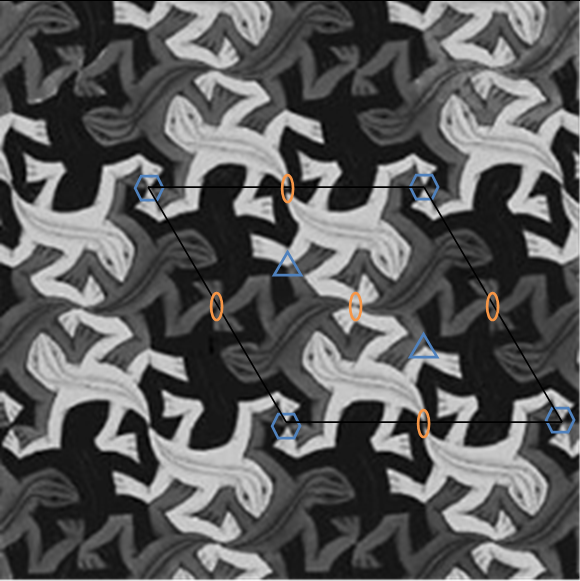}
        \caption{"Reptiles", M. C. Escher, 1943, Fragment, with a $p6|p2$ unit cell superimposed. Symmetry elements depicted in orange are of the proper symmetry, in blue – color-blind. }
        \label{lizards_img}
\end{figure}

A slightly more complex example is given by the "Reptiles" image (Figure ~\ref{lizards_img}). Unlike "Lizards", "Reptiles" have three distinct shades of gray. Therefore, we cannot define the grey group for them anymore, however, we can still obtain a colorblind image for symmetry estimation.
The hexagonal pattern here is easy to recognize by a naked eye. The unit cell overlaid with the image depicts all symmetry features of the picture. Thus, by inspection, the picture possesses $p6$ colorblind symmetry and $p2$ "proper" symmetry, if colors are not ignored. Therefore the full 3-colored (Belov) plane group for this image is $p6|p2$. 
We follow the notations in Ref.~\cite{shubnikov1964colored} for three-color plane groups $G_{col} = G|G'$, where $G$ is the Fedorov plane group and the subgroup $G' \supset G$ of index 3 contains operations that keep the first color fixed. 
Since both Shubnikov (b/w) and Belov (color) plane groups can be represented in terms of the layer groups, where shift along the z axis corresponds to color change, the isomorphic 3D layer group for the lizard pattern is $P6_2$.

\begin{figure}[h!]
\centering
\begin{subfigure}[b]{0.5\textwidth}
    \centering
    \includegraphics[width=\textwidth]{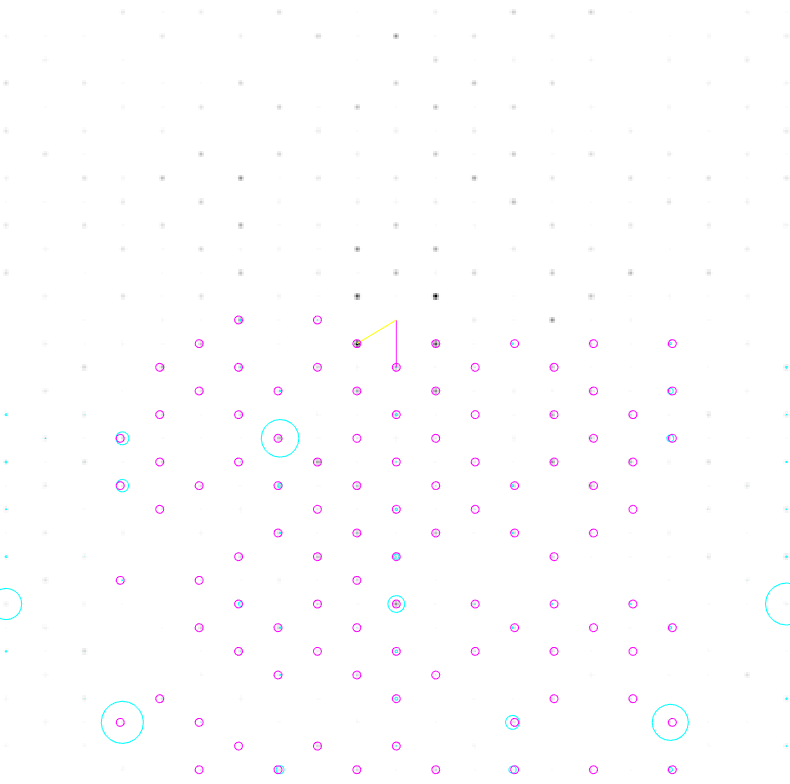}
        \caption{}
        \label{lizards_fft}
\end{subfigure}
\hfill
\begin{subfigure}[b]{0.4\textwidth}
\centering
\includegraphics[width=\textwidth]{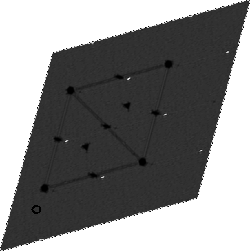}
        \caption{}
        \label{lizards_pom}
\end{subfigure}
\caption{(a) Indexed Fourier Transformation of Figure \ref{lizards_img}. (b) POM for $p6$ refinement overlaid with $p6$ unit cell (the cell is slightly shifted to demonstrate features on the POM)}
\end{figure}

Initial image processing was done in the same way as before. While the first order peaks in the FFT power spectrum still exhibit a hexagonal arrangement, the hexagonal pattern yields to a 2-fold symmetry for the second order FC. (See Figure ~\ref{lizards_fft}) 

On the next step we reconstruct the image and compute phase and amplitude residuals. Groups $p2$ ($\phi_{Res}=20.56$), $p3$ ($\phi_{Res}=18.85$), $p6$ ($\phi_{Res}=22.91$) have phase residuals significantly lower than other groups with consistent cell metrics: $p31m$ ($\phi_{Res}=27.44$), $p3m1$ ($\phi_{Res}=25.72$), $p6mm$ ($\phi_{Res}=39.45$). Only refinement in the $p2$ group gives the correct reconstructed image, therefore we conclude that the Fedorov group for this image is indeed just $p2$. However, the fact that the residuals for $p3$ and $p6$ are very close to $p2$, it suggests the presence of "hidden" symmetry, which in this case is color symmetry. 

In the POM for $p6$ the minima, indicated by the bright spots, correspond non broken 2-axes, and the dark dots correspond to the hidden symmetry elements: 6- and 3-axes. The higher is the order of a symmetry element the more relations between FCs it implies. Therefore 6-axes will appear in the POM darker (i.e. higher residuals) than e.g. 2-axes, since for an origin shift to a six-fold axis a pair of $\left\lbrace\phi_x,\phi_y\right\rbrace$ should obey more restrictions. Thus we conclude that the dark dots on the POM in Figure ~\ref{lizards_pom} are most likely 6-axes. This provides an insight onto the highest symmetry group possible, which is now hidden due to colorization. By applying the Sobel filter and extracting the silhouettes of the lizards, we can see that the image refined in the $p6$ group now clearly resembles the initial one. 

\begin{figure}[h!]
\centering
\includegraphics[width=0.95\textwidth]{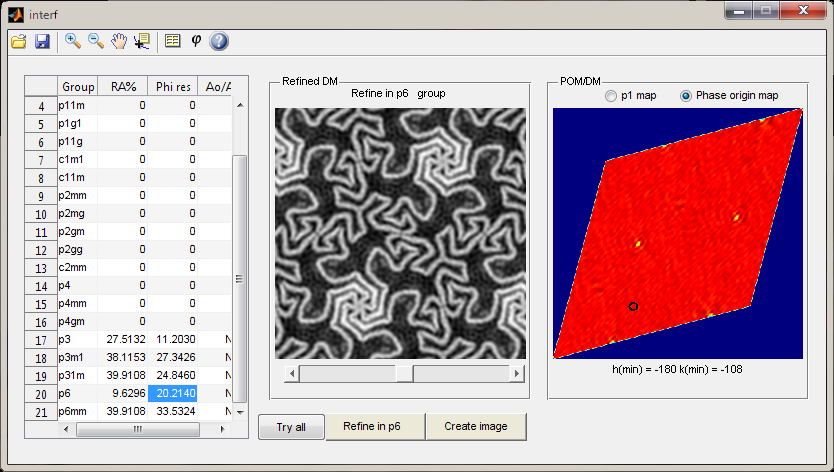}
\caption{Symmetrization of a uniformed silhouette image of Reptiles}
\label{lizards_sym}
\end{figure}

From Figure ~\ref{lizards_sym} one can see that indeed, after colorization is removed, nothing else breaks the highest possible symmetry: $p6$, and the residuals for groups $p6$ and $p3$ (subgroup of $p6$) are lower than they used to be in the colored case. 

An important factor for practical application of the aforementioned techniques of symmetry quantification and refinement is image resolution. If the number of collected FC is quite small (i.e. image resolution is too low), some FC have no mates, and some phase comparisons cannot be performed fully. Usually for high symmetry groups, where the number of related FCs is more than two, only 10-20\% of all FCs have all required mates. Some comparisons may be totally degenerated; in this case there is no difference in the phase statistics between some high- and low- symmetry groups. Therefore, a supergroup will normally have residuals higher than its subgroups. This is indeed the case in this example where the $p3$ group yields lower residuals than the correct $p6$ group. 
\section{Conclusions}

We analyzed the applicability of CIP to symmetry analysis of highly symmetric 2D images. Phase residuals allow us to deduce the most probable plane group, low phase residuals in the groups that don't yield exactly same reconstructed image after refinement  most likely reflect the fact that there are some "broken" or "hidden" symmetry elements. In this case a careful analysis of the phase origin map can often reveal possible candidates for both geometrical and black and white supergroups. An edge detection algorithm, like the Sobel filter can provide some insight about the color-blind supergroup. 

The four examples discussed above illustrate three important features of the CIP for symmetry determination: 
simple identification of the most probable plane group, 
detection of "broken" symmetry elements on an example of a monochrome image, 
detection of "hidden" and "broken" symmetry elements associated with colorization of the initial pattern. 

\pdfoutput=1

\section*{Appendix: Derivation of a numerically efficient version of Eq.\eqref{phi_sym}}

\begin{figure}[h!]
\centering
\begin{subfigure}[b]{0.2\textwidth}
    \centering
    \includegraphics[width=\textwidth]{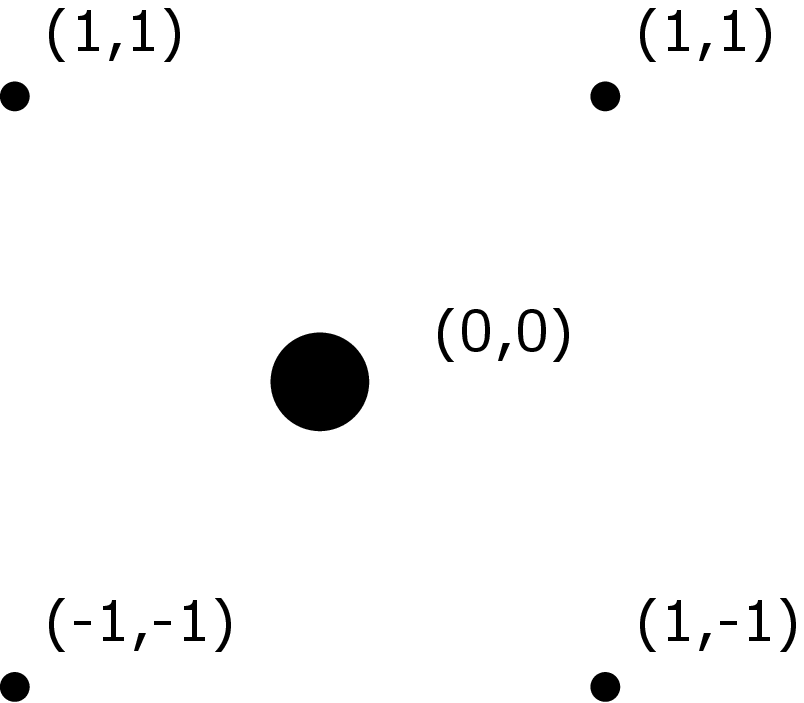}
        \caption{}
        \label{app1}
\end{subfigure}
\hfill
\begin{subfigure}[b]{0.7\textwidth}
\centering
\includegraphics[width=\textwidth]{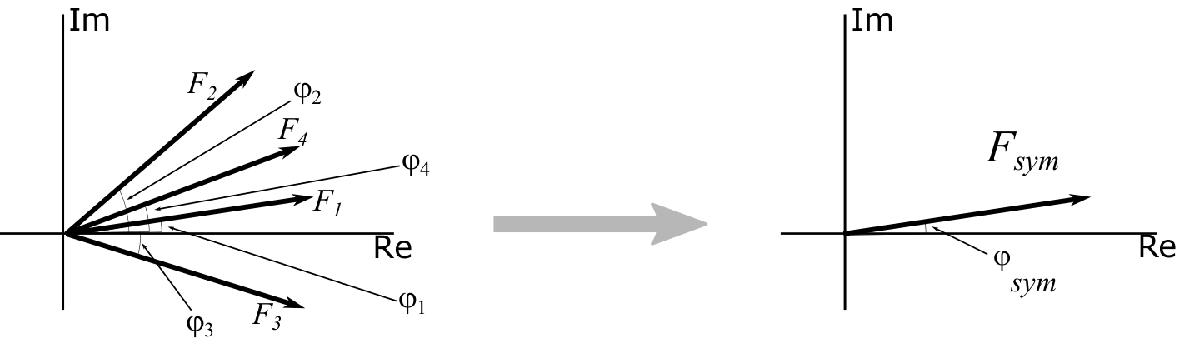}
        \caption{}
        \label{app2}
\end{subfigure}
\caption{(a) Schematic representation of a set of symmetry related FCs. (b) Graphical representation of the symmetrization procedure}
\end{figure}

Original paper by Zou and Hovm\"oller\cite{Zou2006} does not contain derivation of Eq.~\eqref{phi_sym}. However, it can  be obtained by considering strict geometrical relations in the 4-D ($h, k, F, \phi$) Fourier space. Suppose we have a set of symmetry related FCs as determined by a given plane group: (1,1), (-1,1), (-1,-1), (1,-1). By the rules of symmetry, their amplitudes and phases are related as well. The amplitudes of all FCs under consideration should be equal, and  phase increases by $90^{\circ}$ as we go in the counterclockwise direction from the (1,1) FC (See Figure ~\ref{app1}). 

Thus, their complex-valued FCs $\Phi_{hk}=F_{hk}\exp(i\phi_{hk})$ are mutually linked as:
\begin{eqnarray}
\Phi_1=\Phi_{(1,1)}=F\cdot e^{i\phi} \nonumber\\
\Phi_2=\Phi_{(-1,1)}=F\cdot e^{i(\phi+\pi/2)}=i\Phi_1 \nonumber\\
\Phi_3=\Phi_{(-1,-1)}=F\cdot e^{i(\phi+2\pi/2)}=-\Phi_1 \nonumber\\
\Phi_4=\Phi_{(1,-1)}=F\cdot e^{i(\phi+3\pi/2)}=-i\Phi_1 
\label{eq_A1}
\end{eqnarray}

The fore factors $(1, i, -1, -i)$ appear due to the symmetry and do not bear any additional structural information. 
However, if the symmetry is slightly broken, i.e. each amplitude and phase slightly deviates from its ideal value, the equations above transform as:
\begin{eqnarray}
\Phi_1=\Phi_{(1,1)}=F_1\cdot e^{i\phi_1} \nonumber\\
\Phi_2=\Phi_{(-1,1)}=F_2\cdot e^{i(\phi_2+\pi/2)}=iF_2 \cdot  e^{i\phi_2} \nonumber\\
\Phi_3=\Phi_{(-1,-1)}=F_3\cdot e^{i(\phi_3+2\pi/2)}=-F_3 \cdot  e^{i\phi_3} \nonumber\\
\Phi_4=\Phi_{(1,-1)}=F_4 \cdot e^{i(\phi_4+3\pi/3)}=-iF_4 \cdot  e^{i\phi_4} 
\label{eq_A2}
\end{eqnarray}

The amplitudes $(F_1, F_2, F_3, F_4)$ and phases $(\phi_1,\phi_2,\phi_3,\phi_4)$ do not differ significantly, thus, ignoring the symmetry caused fore factors, we can represent the complex-valued FC as a bundle of vectors of almost the same length and pointing in almost the same direction: Figure \ref{app2}. 

The symmetrization procedure replaces this bundle by a single complex number (vector), defined as the complex average of the bundle: $\Phi_{sym}=\sum^n_i\Phi_i \big/ n$. The following equations give the amplitude and the phase of the symmetrized structural factor: 
\begin{equation}
F_{sym}=\frac{\sum_i^n \vert F_i(h,k) \vert}{n}; \quad \phi_{sym}=\arctan\left(\frac{\sum_i^nF_i\sin\phi_i(h,k)}{\sum_i^nF_i\cos\phi_i(h,k)} \right)
\end{equation}

These formulas coincide with Eq.~\eqref{phi_sym}, with the difference, that the weighting factor $w_i$ is now explicitly expressed in terms of the amplitude. Also, the origin of the $s$ factor is now explicitly revealed as the presence of glide lines changes the fore-factor in Eqs.~\eqref{eq_A1},~\eqref{eq_A2} by rotating the corresponding vector either $90^{\circ}$ or $180^{\circ}$ degrees, and thus it has to be rotated back for correct averaging. Complex FCs used for image reconstruction are now given as:
\begin{equation}
\Phi_1=\Phi_{sym};\quad \Phi_2=-\Phi_{sym};\quad \Phi_3=i\Phi_{sym};\quad \Phi_4=-i\Phi_{sym};
\end{equation}
From the numerical point of view, formula~\eqref{phi_sym} is not very efficient, since it requires serial calculation of $2n+1$ trigonometric functions, which is a time-consuming routine. As it is shown here, Eq.~\eqref{phi_sym} computes the argument of a sum of many complex numbers, it is worth keeping coefficients of the Fourier transformation for the image in the complex-valued form (i.e. not to separate phase and amplitude). Then the Eq.~\eqref{phi_sym} can be rewritten as: $\phi_{sym}=\arg\left(\sum_is_i\Phi_i\right)$ , which uses only one trigonometric function instead. Using the latter formula instead of ~\eqref{phi_sym} yields two to four times (depends on the number of relations) boost in the performance. 

\bibliographystyle{unsrt}  
\bibliography{bibliography}
\end{document}